\documentclass[prc,aps,znofootinbib,showkeys,showpacs,twocolumn,floatfix]{revtex4} 
\usepackage{epsfig}
\usepackage{graphicx}
\usepackage{ulem}

\begin{document}
\title{Pairing and  specific heat in hot nuclei }
\author{ Danilo Gambacurta}
\email{gambacurta@ganil.fr}
\affiliation{GANIL, CEA and CNRS/IN2P3, Bo\^ite Postale 55027, 14076 Caen Cedex, France}
% \author{Denis Lacroix} 
% \email{lacroix@ganil.fr}
% \affiliation{GANIL, CEA and CNRS/IN2P3, Bo\^ite Postale 55027, 14076 Caen Cedex, France}
\author{Denis Lacroix} \email{lacroix@ganil.fr}
\affiliation{Institut de Physique Nucl\'eaire, IN2P3-CNRS, Universit\'e 
Paris-Sud, F-91406 Orsay Cedex, France}
\affiliation{GANIL, CEA/DSM and CNRS/IN2P3, Bo\^ite Postale 55027, 14076 
Caen Cedex, France}

\author{N. Sandulescu} 
\email{sandulescu@theory.nipne.ro}
\affiliation{National Institute of Physics and Nuclear Engineering, P.O.Box MG-6, Magurele-Bucharest, Romania}

\def\be{\begin{equation}}
\def\ee{\end{equation}}
\def\Fe{$^{56}$Fe}
\def\Cr{$^{66}$Cr}
\def\Tr{{\rm Tr}}

\begin{abstract}
The thermodynamics of pairing phase-transition in nuclei is studied in the
canonical ensemble and treating the pairing correlations in a finite-temperature
 variation after projection BCS approach (FT-VAP). Due to the restoration of
particle number conservation, the pairing gap and the specific heat
calculated in the FT-VAP approach vary smoothly with the temperature, indicating
a gradual transition from the superfluid to the normal phase, as expected in
finite systems. We have checked that the predictions of the FT-VAP approach are
very accurate when compared to the results obtained by an exact diagonalization 
of the pairing Hamiltonian. The influence of pairing correlations on specific heat is
analysed for the isotopes $^{161,162}$Dy and $^{171,172}$Yb.  
It is shown that
the FT-VAP approach, applied with a level density provided by mean field
calculations and supplemented, at high energies, by the level density of the
back-shifted Fermi gas model, can approximate reasonably well the main 
properties of specific heat extracted from experimental data. However, the 
detailed shape of the calculated specific heat is rather sensitive to the
assumption made for the mean field.

\end{abstract}

\pacs{25.60.Je,25.40.Hs,24.30.Cz,21.10.Re,21.60.Jz}
\keywords{pairing, thermodynamics, nuclear models}
\maketitle

\section{Introduction}

In the last decade there is a renewed  interest for studying
the fingerprints of pairing correlations in the thermodynamic 
properties of excited nuclei. This interest was triggered by 
the new accurate measurements of level density at low excitation 
energies. Thus, a special attention was payed lately to the influence of 
pairing on the low-temperature behavior of the specific heat in 
the isotopes $^{161,162}$Dy and $^{171,172}$Yb, extracted from the
level density measurements performed by the Oslo group \cite{melby99}. 
The possible thermal signatures of pairing correlations in these rare
earth isotopes have been studied either with schematic models or 
employing various approximations which go beyond the standard BCS
approach. As it is by now well documented,
due to its drawbacks, i.e., particle number fluctuation and 
quasiparticle parity mixing, the BCS theory is not well-suited
to describe pairing effects in hot nuclei. One alternative to cure these 
drawbacks is to use the particle-number projected BCS approximation extended to 
finite temperature. How this approximation can be implemented for performing
variation after projection  calculations at finite temperature (FT-VAP)
was recently discussed in Ref \cite{Gam12}. This approach will be applied here
to investigate the effect of pairing on thermal properties of
Dy and Yb isotopes mentioned above. 

In Ref. \cite{schiller01} it  was argued that the S-shaped form of the 
specific heat in Dy and Yb isotopes is generated by the transition from 
the superfluid to the normal phase. This conclusion was drawn by the 
comparison with the specific heat of a 
non-interacting Fermi gas which was described by the 
back-shifted Bethe formula \cite{gc65}. However, this comparison 
is misleading because the Bethe formula is not valid for low energy 
excitations (e.g., see  \cite{bm69}), where the pairing correlations 
are expected to be important. In this study we shall re-analyze this issue 
in the framework of FT-VAP approximation and using for low energy excitations 
level densities extracted from self-consistent mean field calculations.

The article has the following structure. First, we present the calculation scheme we use
to evaluate the thermodynamic properties of nuclei related to the pairing interaction.
Afterwards,  using single-particle spectra generated by self-consistent mean field models, 
we analyze the effect of pairing on the partition function and  heat capacity. 
Finally, a critical discussion on the comparison with experimental data is made.   

\section{Statistical treatment of pairing interaction in finite systems}
\label{sec:protocol}

To analyze the effects of pairing correlations upon the thermodynamic properties
of hot nuclei we consider the  hamiltonian:
\begin{eqnarray}
\hat H= \sum_i \epsilon_i a_i^+a_i + \sum_{i,j} V_{ij}  a^+_i a^+_{\bar{i}} a_{\bar{j}} a_j \label{hamN},
\end{eqnarray}
where the second term is the pairing interaction which scatters pairs among time-reversed single-particle
states $(i,\bar i)$. This hamiltonian can be used for realistic description of pairing correlations in
heavy nuclei for which proton-neutron pairing can be neglected.

For many years the pairing correlations in hot nuclei have been described in the framework of 
finite-temperature BCS/HFB models (FT-BCS/HFB) \cite{FTHFB}. However, FT-BCS is the
proper theory for infinite systems but not for finite systems such as atomic nuclei. 
This is reflected in the fact that FT-BCS predicts a sharp (second order) superfluid-normal 
phase transition, in contrast to a smooth transition, as expected in hot nuclei \cite{Gam12}. 
This drawback of FT-BCS is related to the improper treatment of particle number conservation, pleading in favor of approaches that explicitly conserve particle number.
 Several calculation schemes have been proposed along this line, e.g., based on Shell-Model 
Monte-Carlo (SMMC) calculations \cite{Alh03} or the method proposed recently in Ref. \cite{Qua10}.  
In the present study we shall explore a variational method which is the generalization of the
particle-number projected BCS model to finite temperature. This method, called below 
finite-temperature variation after projection (FT-VAP) approach, has been recently  
tested \cite{Ese93,Gam12} in the case of schematic pairing models. Here the FT-VAP approach
will be extended and applied to describe thermodynamic properties of hot nuclei. In order to 
test the validity of FT-VAP in realistic applications, we perform also thermodynamic
calculations based on the spectrum of the pairing Hamiltonian (\ref{hamN}) obtained by
 direct diagonalization.
   
\subsection{Input for pairing Hamiltonian}

In the applications presented below the single-particle (s.p.) states employed in the 
Hamiltonian (\ref{hamN}) are taken from mean-field models. Thus, we have used two sets of s.p. 
states obtained, respectively, from Skyrme-Hartree-Fock (HF) and Relativistic 
Mean-Field (RMF) calculations allowing, in both cases, axial symmetry deformation.
The Skyrme-HF equations are solved with the code EV8 \cite{Bon05} and using the force 
Sly4 \cite{sly4} while the RMF calculations are done with the force PK1  \cite{Pk1}.
The s.p. energies around the Fermi energy  for the isotopes $^{162}$Dy 
and $^{172}$Yb obtained in Skyrme-HF and RMF \cite{liu} calculations
are shown in Figure 1. These energies are also considered in the
calculations done for the odd isotopes $^{161}$Dy and $^{171}$Yb.
In HF and RMF calculations  a deformation of about $\beta_2=0.35$ is found both 
for $^{162}$Dy and $^{172}$Yb, $\beta_2$ being defined by
\begin{equation}
\beta_2= \Big(\frac{5\pi}{9} \Big)^\frac{1}{2} \frac{\langle \hat Q_2 \rangle}{A R_0^2},
\label{Eq:beta}
\end{equation} 
where $A$ denotes the mass number, $R_0=1.2 A^{1/3}$ and $\hat Q_2$ is the quadrupole operator.

\begin{figure}[htbp] 
\includegraphics[width=9cm]{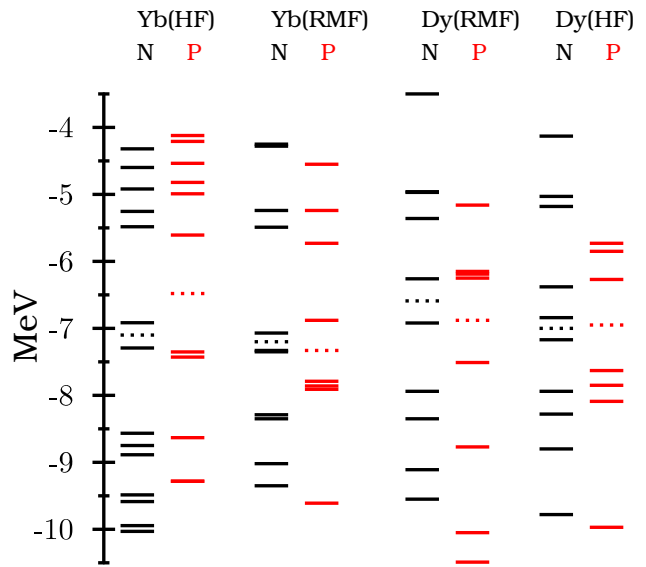} 
\caption{ (color online) Neutrons (N) and protons (P) single particle energies 
around the Fermi energies (dotted lines) for $^{162}$Dy and $^{172}$Yb obtained from 
HF-Sly4 and RMF-PK1 calculations \cite{liu}.}
\label{Fig:sps} 
\end{figure} 

 As commonly done in BCS calculations at zero temperature, pairing interaction is
 considered acting in a limited energy window around the Fermi level. In this window
 the matrix elements of the pairing interaction are taken equal to a constant strength,
 i.e., $V_{ij}=G$. In the calculations presented below for Dy and Yb isotopes we have
 considered an energy window of 3 MeV. The pairing strength $G$ is fixed at BCS
 level in order to give a pairing gap approximately equal to $ 0.8$ MeV both for protons
 and neutrons.  The number of single  particle states and the  active nucleons as well as
 the corresponding pairing strength $G$ for the energy window of 3 MeV are given in Table    
 \ref{tab:input}.

\begin {table} 
 
\begin{tabular}{|ccccc|}
 \hline
 &&~~~~~~~~~~~~~~~Neutron& &\\
 \hline
 &Yb (HF)&Yb (RMF)&Dy (RMF)& Dy(HF)\\
\hline 
$N_{sp}$  &14&  11     & 10&10 \\
 $n_{a}$  &16& 12     &10&10   \\
 $G (MeV)$&0.260&  0.270&0.320&0.284\\
\hline
\hline
& &~~~~~~~~~~~~~~~Proton& &\\
 \hline
&Yb (HF) &Yb (RMF)&Dy (RMF)& Dy(HF)\\
\hline 
$N_{sp}$  &11& 8     & 8&9 \\
 $n_{a}$  &10& 8     &8&8   \\
 $G (MeV)$&0.332&  0.345&0.385&0.327\\
\hline
\end{tabular}
 \caption {Number of single  particle levels ($N_{sp}$) and active nucleons ($n_{a}$) 
in the active 3 MeV window
 around the Fermi energy for neutrons (upper table)  and protons (lower table). 
The strength $G$ is  fixed in order to have a 
pairing BCS gap of 0.8 MeV for both the neutrons and protons.}\label{tab:input}
\end{table}

\subsection{Pairing treatment in canonical ensemble }

 The thermodynamic properties of pairing interaction are calculated here in
 the canonical ensemble \cite{Sum07}. The key quantity is the canonical 
 partition function
\begin{equation}
Z(\beta) = \Tr e^{-\beta \hat H} \equiv {\rm Tr}(D_N[\beta]) , \label{excitation-partition}
\end{equation}
where $\beta= 1/T$ and  $D_N[\beta]$  denotes the statistical  $N$-body density operator. 
The trace has to be taken 
over the states with well-defined proton and neutron numbers.
Using the eigenstates $\{ E_n\}$ of the hamiltonian $H$, the partition function can 
be written as
\footnote{Usually the energy are measured with respect to the ground state energy $E_0$, it is often convenient to 
introduce the so-called excitation partition function $Z'\equiv Ze^{\beta E_0}.$ \cite{Alh03}
Then, $E_n$ correspond to the energy with respect to the ground state energy.}

\begin{equation}
Z =  \sum_n \rho(E_n)  e^{-\beta E_n} \label{eq:Zexact}
\end{equation} 
where $\rho(E_n)$ stands for the level density. Quantities 
of physical interest are obtained from the derivatives of the partition function. 
For instance, the thermal energy can be evaluated through
\begin{eqnarray}
 E(T)  & = & - \frac{ \partial \ln Z}{ \partial \beta} =  \langle H \rangle
\nonumber \\
&=& Z^{-1} \sum_n \rho(E_n) E_n e^{-\beta E_n},
\end{eqnarray}
where $ \langle X \rangle = {\rm Tr}(X ~D_N[\beta]) $. 

\begin{figure}[htbp] 
\includegraphics[width=8cm]{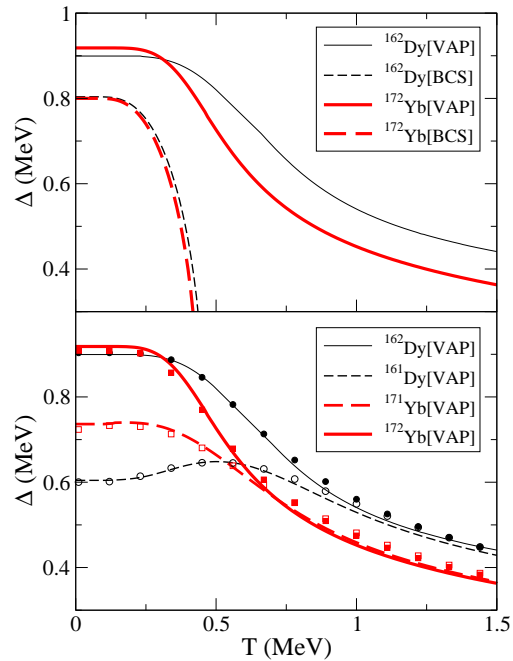} 
\caption{ (color online) Top: neutron pairing gap as a function of 
temperature for $^{162}$Dy and  $^{172}$Yb obtained in  FT-VAP and
FT-BCS approximations.
 Bottom: comparison between the pairing gaps in odd-even and even-even 
 Dy and Yb isotopes obtained in FT-VAP approximation. By symbols are shown
 the gaps corresponding to exact diagonalization, calculated with 
 formula (\ref{eq:gapexact}).  }. 
\label{Fig:allgaps} 
\end{figure}

In the case of the Hamiltonian (1), using a pairing force restricted to a narrow window
around the Fermi level, one can obtain the eigen-energies $E_n$ by direct diagonalization
in subspaces of given seniority.
Such approach is certainly the most direct way to
describe  statistical properties of pairing at fixed particle number. 
However, this approach can be only applied at low temperatures because only
in this case the pairing active window can be taken small enough for allowing
exact diagonalization. Alternatively, the statistical properties of
pairing interaction are usually investigated in the framework of quasiparticle
models, such as BCS or HFB, in which the excitations can be easily obtained 
and the pairing gap is built  from outset. Thus, in the finite-temperature
BCS (FT-BCS) approximation the partition function (\ref{excitation-partition}) is
calculated with the BCS effective Hamiltonian while the variation of the pairing 
correlations with the temperature are characterized by 
the pairing gap (see for example \cite{Goodman}). 
For illustration, in the top panel of 
Fig. \ref{Fig:allgaps} it is shown how the pairing gap varies with
the temperature in the isotopes $^{162}$Dy and $^{172}$Yb. 
The results corresponds to the  RMF s.p. spectrum.
As can be seen, the FT-BCS predicts a sharp second order type transition 
from the superfluid to the normal phase, at variance with what it is expected 
for small finite systems. As it is well known, the  sharp transition 
predicted by FT-BCS is connected to the breaking of particle number 
conservation, leading automatically to a grand-canonical treatment instead 
of a canonical one. 

A more involved but more appropriate approach, which keeps the physical insight of 
quasiparticle models, can be obtained working with the particle number projected 
density 
\begin{eqnarray} 
\hat D_N = \frac{1}{Z} {\hat P}_N
\exp(-\beta \hat h)  {\hat P}_N , \label{eq:dn} 
\end{eqnarray} 
where $Z={\rm Tr}({\hat P}_N \exp(-\beta \hat h)  {\hat P}_N)$,  $\hat h$ is 
the BCS effective Hamiltonian, and  ${\hat P}_N$ is the projector onto good particle number.
This density is used below in a variation after projection (VAP) scheme to minimize 
the Helmholtz free energy. This approach has been proposed already some times 
ago \cite{Ese93} and recently tested in the Richardson model \cite{Gam12}.  
Technical details related to the solution of the FT-VAP equations can be 
found in the latter reference. 

An illustration of the pairing gap obtained using the FT-VAP approach can be seen in the
top panel of Fig. \ref{Fig:allgaps} where are shown the results for the isotopes
$^{162}$Dy and $^{172}$Yb. Contrary to the FT-BCS approximation, FT-VAP predicts a smooth
transition from the superfluid to to the normal phase, with a non-vanishing pairing gap
extended up to high temperatures.  To test the predictions of FT-VAP we have also done
calculations using the exact solutions of the pairing Hamiltonian obtained by direct
diagonalization and, similarly to ref.  \cite{Gam12}, we have estimated the pairing gap
by the formula
\begin{eqnarray}
\Delta=\sqrt{-G (E-E_{0}}) \label{eq:gapexact}
\end{eqnarray}
where $E$ is the total exact energy and $E_0$ is given by  
\begin{eqnarray}
E_{0}=\sum_i \left(\varepsilon_i-\frac{G}{2} n_i\right)n_i.
\end{eqnarray} 
The s.p. occupation numbers $n_i$ are those deduced from the exact calculation. 
The gap obtained by the exact diagonalization and the FT-VAP are compared in the
bottom part of Fig. \ref{Fig:allgaps}. As can be seen, the two calculations give
similar gaps, indicating the high  accuracy of the FT-VAP approach.
In Fig. \ref{Fig:allgaps} are given also the pairing gaps obtained for the odd nuclei.
As expected, at low $T$, the gaps in odd systems are smaller than in the even systems.
This difference tends to disappear as the temperature increases. For  $^{161}$Dy and to
a lesser extend for  $^{171}$Yb, we also observe a small re-entrance effect 
around $T=0.5$ MeV (resp. around 0.2-0.3 MeV). This effect is related to the thermal
scattering of the odd particle in high energy s.p. levels, which increases the available
phase space for the pair scattering.

\section{Heat capacity}

\begin{figure}[htbp] 
\includegraphics[width=8cm]{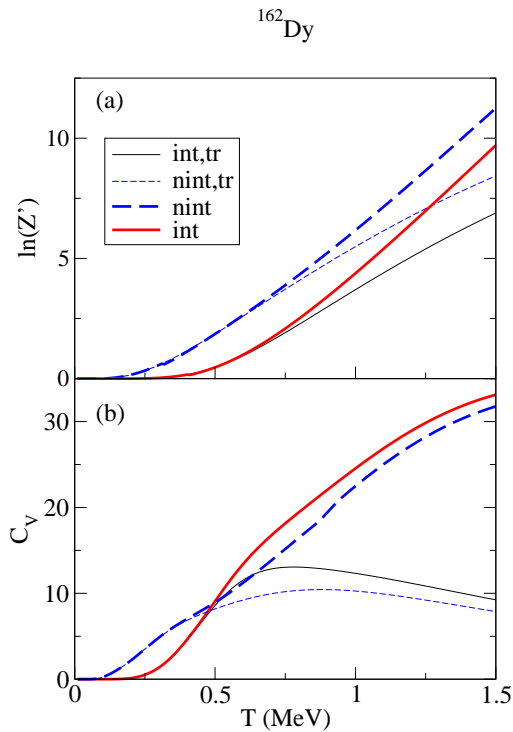} 
\caption{ (color online) 
Top: Partition functions for $^{162}$Dy. The labels "nint-tr" and
"int-tr"  correspond  to the partition function obtained, respectively,
without and with the pairing interaction  and considering the single-particles
states from a window of 3 MeV. The label "nint"  corresponds to the case of
non-interacting particles in the window of 7 MeV  while the results for the
extended partition function $\ln Z'$ (\ref{extended-partition}) are 
labeled by "int". 
Bottom: Corresponding heat capacities. }
\label{fig:lnz} 
\end{figure}

Statistical properties of hot nuclei are 
commonly described by the heat capacity extracted from experimental
level density. The specific heat is calculated from the second derivative 
of the partition function
\begin{eqnarray}
C_V(T) &=& \beta^2 \frac{\partial^2 \ln Z }{\partial^2 \beta}.
\label{eq:cv}
\end{eqnarray}
The partition function and the specific heat for $^{162}$Dy, obtained
with the s.p. spectrum truncated by a 3 MeV energy window around the 
Fermi level, are shown in Fig. \ref{fig:lnz}  by the lines labeled "int,tr". For
comparison are also shown, by the lines labeled "nint,tr", the
results corresponding to the free spectrum (G=0). It can be seen
that, in both calculations, the specific heat has an unphysical behaviour
at high temperatures. To cure this behaviour one should considering  s.p. states 
from energy windows around the Fermi energy larger than 3 MeV. Since the pairing 
force is of zero range, in principle the effect of pairing correlations can be 
study with any energy window provided the strength of the force is 
adjusted such as to preserve the amount of pairing correlations, measured, 
for instance, by the BCS pairing gap. However, this strategy cannot be applied for 
too large energy windows because the number of  excited states  which 
can be built on from the s.p. levels becomes too large to be handled  in the
FT-VAP or direct diagonalization calculations.  The calculations can be, however, simplified by
taking into account the fact that the  pairing correlations have little effect on the excitations 
of high energies. Based on this  argument, we adopt here the calculation scheme  proposed in 
Ref. \cite{Alh03} and assume that the pairing interaction affects the excitation energies generated
by the s.p. energies taken from a restricted region around the Fermi energy while the rest of
s.p. spectrum is treated as in the case of non-interacting particles. For the partition function
this approximation can be written as \cite{Alh03}:
\begin{eqnarray}
\ln Z'_{int} = \ln Z'_{int,tr} -\ln Z'_{nint,tr} +\ln Z'_{nint} \;,
\label{extended-partition}
\end{eqnarray}
where $Z'_{int,tr}$ and $Z'_{nint,tr}$ are, respectively, the partition function
calculated with and without pairing interactions in the truncated space, here
the 3 MeV window. $Z'_{nint}$ is the partition function calculated without the
pairing interaction considering all the s.p. levels from a  window of 7 MeV around 
the Fermi energy. We have  checked that enlarging further this space does not affect 
the results in the temperature region considered below.

  As an illustration of the use of formula (\ref{extended-partition}), in top panel of figure \ref{fig:lnz}
 are shown the various  contributions to the partition function for the case of  $^{162}$Dy.
 The results obtained with the extended partition function are labeled by "int". 
 For temperatures below $T \simeq 0.5$ MeV the  canonical partition function for the non-interacting 
 case (G=0) is evaluated by a direct counting of the excited states. For higher temperatures, when
 this procedure becomes numerically difficult, the canonical partition function $Z'_{nint}$ is evaluated,
 through the saddle point approximation, from the partition function in the grand-canonical ensemble.
 This approximation is discussed in Ref.  \cite{Alh03}  and the main formulas are given in 
 appendix \ref{app:gc}.  It should be mentioned that this approximation does not work well at low 
 temperatures, which is the reason why  we apply it here for temperatures above $T \simeq 0.5$ MeV.

From the partition function calculated as explained above we have obtained the heat capacity
by applying Eq. (\ref{eq:cv}).
The results for $^{162}$Dy  are shown  in the bottom
panel of Fig. \ref{fig:lnz}. This figure clearly shows that the calculations done in the truncated 
space become unreliable at higher temperatures. Indeed, from Fig. \ref{fig:lnz} one can notice
that for temperatures larger than  $T \simeq 0.5$ MeV the heat capacity calculated in the truncated 
space becomes much smaller than the specific heat evaluated with the extended partition 
function (\ref{extended-partition}), which reflects the important contribution of high energy 
excitations which are artificially cut in the former case.  On the other hand, as expected from the 
assumption we have made relative to the calculation of the extended partition function, 
it can be seen that the specific heats obtained with and without the  pairing interaction  
become similar at high temperatures.  

 The most important information which can be extracted from  Fig. \ref{fig:lnz} is the effect of 
 pairing correlations on the specific heat. Comparing the results obtained with and without the
 pairing force, one can thus see that below  $T \simeq 0.5$ MeV  the specific heat is strongly 
 suppressed by pairing correlations, an effect which can be traced back to the large pairing gap
 in the low  temperature region (see Fig. \ref{Fig:allgaps}). Above  $T \simeq 0.5$ MeV,  the specific heat becomes
 larger than the results for non-interacting particles and then, at much higher temperatures, it
 goes closer to the latter. This dependence of the specific heat on temperature is commonly
 referred to as a "S-shape" behaviour. 

 We have also analyzed the effect of pairing on specific heat separately for protons and neutrons.
 The results for the isotopes $^{161,162}$Dy and $^{171,172}$Yb are given in 
 Figs. \ref{fig:incdelta}  and \ref{fig:vargap}. 
 In these figures are shown the results obtained by applying,
 in the truncated space, the FT-VAP approximation and the method of direct diagonalization.
 It can be seen that both treatments give similar results, confirming again the good predictive 
 power of the FT-VAP approximation. From these figures we can observe that for even number of
 neutrons or protons there is a large  difference between the non-interacting and interacting
 case, as noticed for the total specific heat of $^{162}$Dy discussed above. 
 However, for odd  number of neutrons this difference is much smaller. Consequently, 
 the effect of pairing on total specific heat in even-even 
 and odd-even isotopes is  different. 
 This is illustrated in  Figure \ref{fig:odd-even}.
 As expected, at large temperatures, for which the pairing correlations are not anymore effective,
 the difference between even-even and odd-even systems is becoming very small.

\begin{figure}[htbp] 
% \vspace{0.8cm}
\includegraphics[width=8cm]{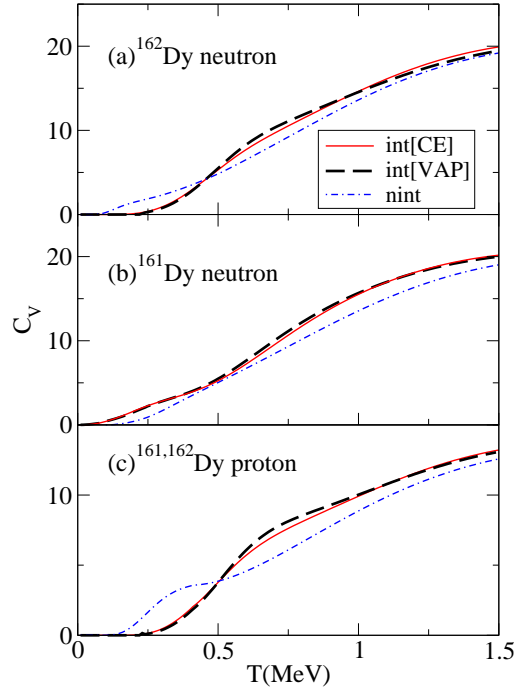} 
\caption{ (color online) Neutron specific heat for $^{162}$Dy (top), $^{161}$Dy (middle) and
the proton specific heat for $^{162,161}$Dy (bottom) .
The lines labeled by "int(VAP)" and "int(CE)" give the results corresponding to FT-VAP and, respectively,
to exact diagonalization while "nint" are the results for non-interacting particles. The 
calculations are done with the extended partition function (\ref{extended-partition}).}
\label{fig:Dycv} 
\end{figure} 
\begin{figure}[htbp] 
% \vspace{0.8cm}
\includegraphics[width=8cm]{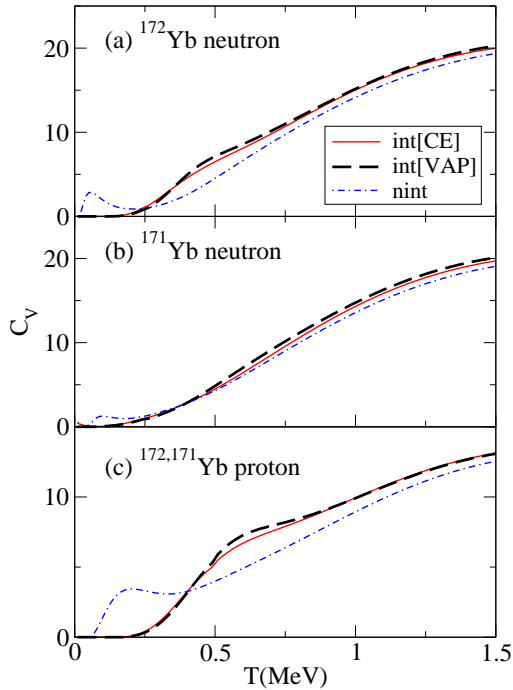} 
\caption{ (color online) The same as in  Fig. \ref{fig:Dycv} but for Yb isotopes.} 
\label{fig:Ybcv} 
\end{figure}
\begin{figure}[htbp] 
% \vspace{0.8cm}
\includegraphics[width=8cm]{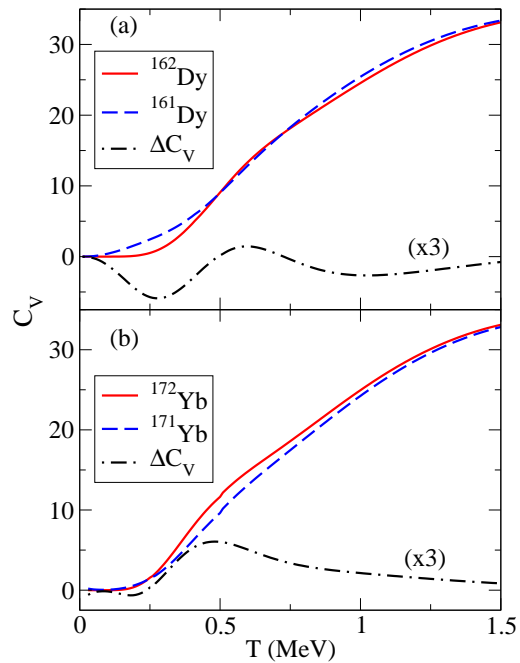} 
\caption{ (color online) Total (neutron+proton) heat capacity in even-even and odd-even Dy (top)
and Yb (bottom) isotopes. The difference $\Delta C_V = C_V^{even} - C_V^{odd}$ is also shown 
by dot-dashed line (note that the difference is multiplied by a factor $3$).} 
\label{fig:odd-even} 
\end{figure}

\begin{figure}[htbp] 
% \vspace{0.8cm}
\includegraphics[width=8cm]{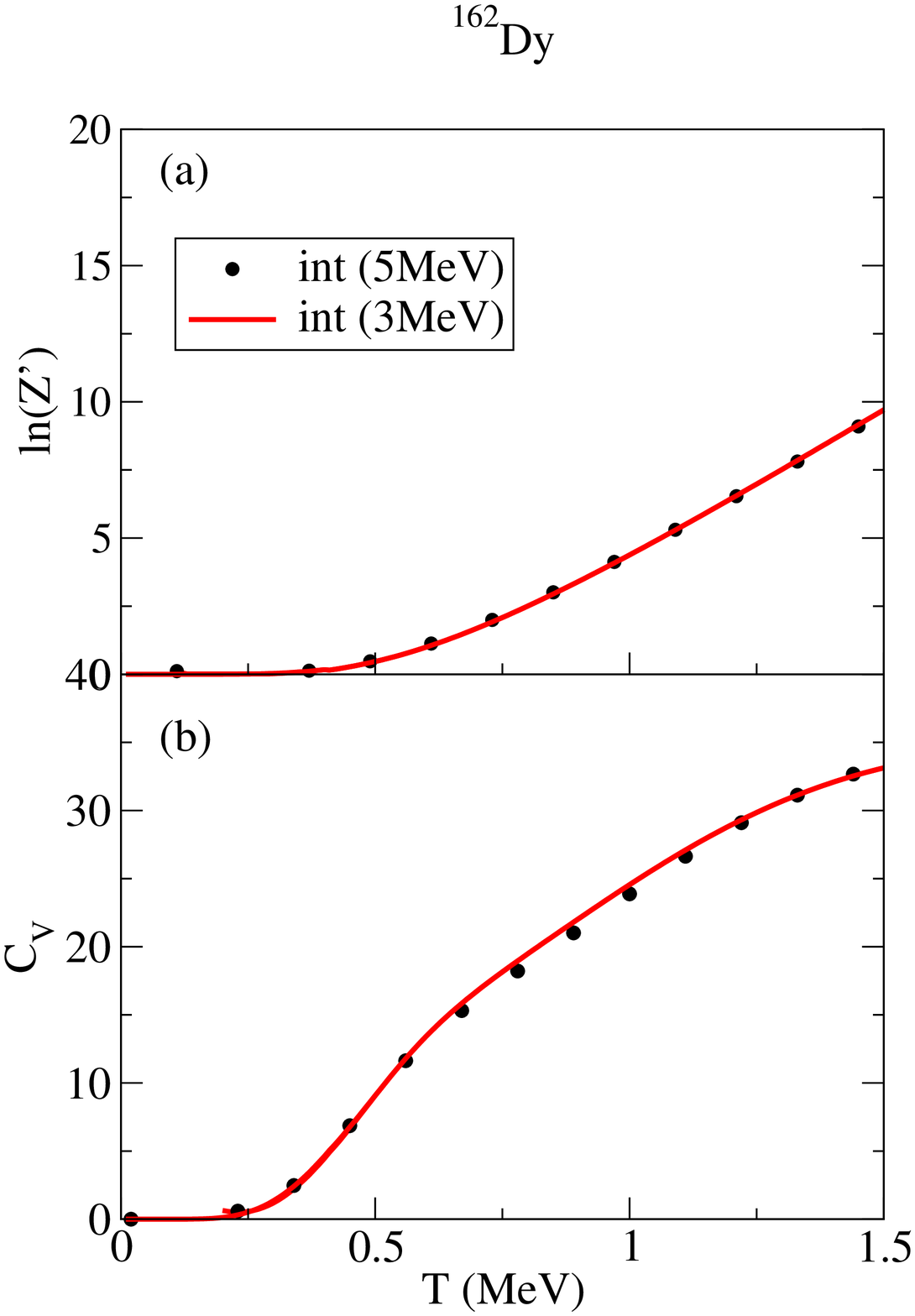} 
\caption{ (color online) Comparison between the partition functions
and the heat capacities in $^{162}$Dy obtained with the pairing 
interaction acting
in two energy windows, of 3 MeV and 5 MeV. The calculations are done
with the extended partition function (10) and fixing the pairing strength
such as to get the same BCS pairing gap in the two energy windows.}
\label{fig:incdelta} 
\end{figure}

To check the convergence of the results with the energy window chosen for the pairing
interaction, we have  performed a second calculations by enlarging the interacting  
window from 3 MeV to  5 MeV. In both calculations the strength of the pairing force
was fixed in order to get a BCS gap equal to $\Delta =0.8$ MeV and the extended partition
function was calculated with a window of 7 MeV. The results for the specific 
heat obtained for the two energy windows are shown in Fig. \ref{fig:incdelta}. 
It can be  seen that the two calculations  are on top of each other.

The specific heat calculated in the present approach depends on the assumptions made on
single-particle energies and the strength of the pairing interaction employed in 
the Hamiltonian (\ref{hamN}).

The sensitivity of the specific heat to the strength of the pairing interaction is illustrated
in Fig. \ref{fig:vargap} where it is  shown the specific heat of neutrons in  $^{162}$Dy  
obtained with three values for the strength corresponding to the  
BCS gaps $\Delta =\{0.6, 0.8, 1.0\}$ MeV. As expected, the S-shape is becoming more pronounced 
when the strength of the force is increasing.

\begin{figure}[htbp] 
% \vspace{0.8cm}
\includegraphics[width=8cm]{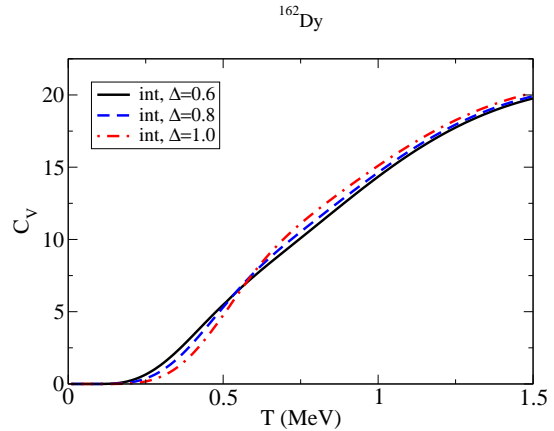} 
\caption{ (color online) 
Heat capacity in $^{162}$Dy for three different strengths of pairing interaction
corresponding to the BCS gaps $\Delta =\{0.6, 0.8, 1.0 \}$ MeV. }
\label{fig:vargap}
\end{figure} 

\begin{figure}[htbp] 
% \vspace{0.8cm}
\includegraphics[width=8cm]{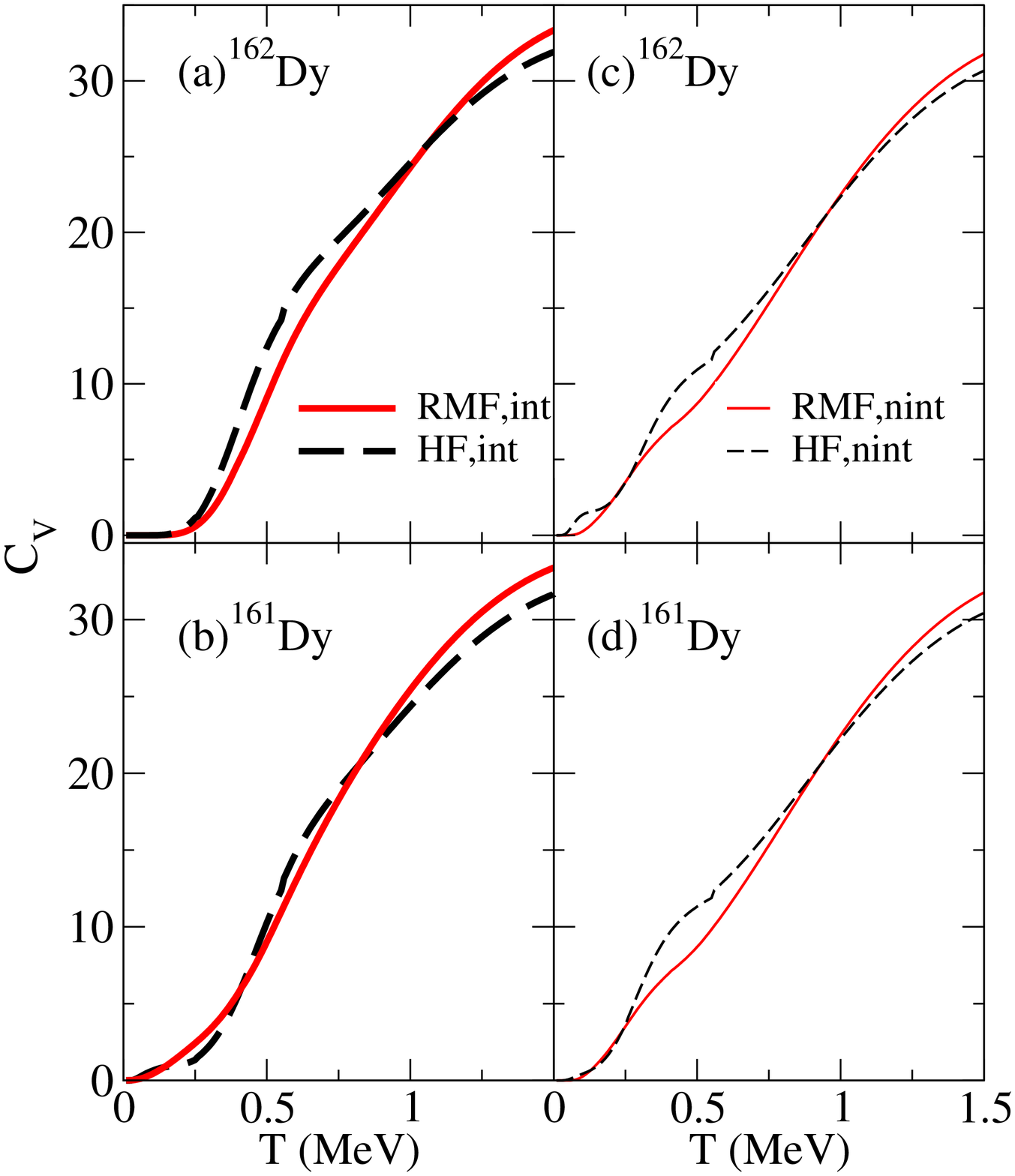} 
\caption{ (color online) Heat capacities in $^{162}$Dy and $^{161}$Dy obtained using two 
different sets of single-particle energies corresponding to RMF and  HF mean fields.
In the left (right)  panels are shown the results with (without) the  pairing interaction
included. The results with the pairing interaction correspond to a BCS gap of 0.8 MeV.}
\label{fig:cvsp} 
\end{figure}

\begin{figure}[htbp] 
% \vspace{0.8cm}
\includegraphics[width=8cm]{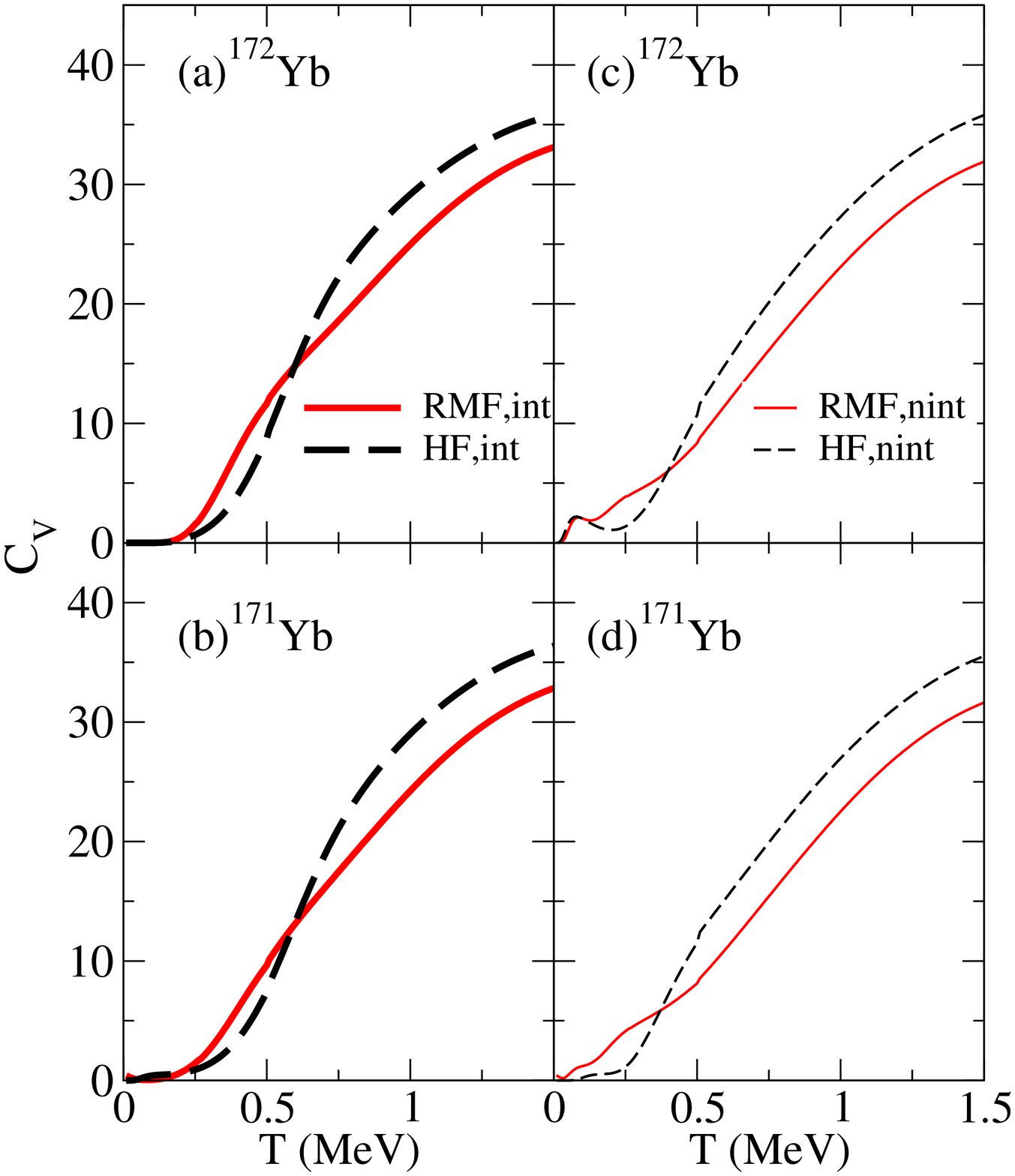} 
\caption{ (color online)The same as  in Fig. \ref{fig:cvsp} but for $^{172}$Yb and $^{171}$Yb.
} 
\label{fig:cvsp2} 
\end{figure}

How much depends the specific heat  on the single-particle spectrum can be seen from  Figs \ref{fig:cvsp} 
and \ref{fig:cvsp2}, where are compared the results obtained with the energies provided by the Skyrme-HF
and the RMF calculations. In both cases the strength of the pairing force is adjusted to obtain the same
BCS gap. It can be seen that there are significant differences between the results obtained with the
two mean fields, especially for Yb isotopes. These differences are related to the distribution of
the single particle levels around the Fermi energy, shown in Fig. 1. Consequently, a change of the 
pairing strength can be easily compensated by a change in the single particle energies,
which makes the comparison with experimental data rather difficult, especially keeping in 
mind the current debate regarding the possibility for an energy density functional approach 
to be predictive for the effective single-particle energies \cite{Duguet2012}.

As it has been observed in previous studies (e.g., see \cite{Nakada05}), the presence of a bump or a S-shape in
the heat capacity does not necessarily signs the transition from superfluid to normal system. This is clearly proved 
by the non-interacting case, where bumps are sometimes observed in the low temperature regime. 
Such a behavior is indeed expected if isolated single-particle states are lying very close to the Fermi energy 
and a gap in energy exist between these states and other surrounding states. This situation happens for instance in 
the case of Yb nuclei for neutrons (see Fig. \ref{Fig:sps} (left)) and is reflected by a pronounced bumps at low 
temperature (Figure \ref{fig:Ybcv}). 

To be more quantitative, let us consider the schematic situation of a set of $N$ degenerated 
two-level system centered around the Fermi energy. The system is assumed to be isolated from the other states. 
Then, the non-interacting hamiltonian can be written as:
\begin{eqnarray}
H &=& \sum_{k=1,N} \frac{\Delta  \varepsilon  }{2} (a^\dagger_{2,k} a_{2,k} - a^\dagger_{1,k} a_{1,k} ).
 \end{eqnarray} 
For this system, the canonical partition function writes:
\begin{eqnarray}
Z(\beta) &=& 2^N \left[ \cosh\left(  \beta \frac{ \Delta  \varepsilon  }{2} \right) \right]^N,
\end{eqnarray}    
leading to:
\begin{eqnarray}
C_V & = & N \left( \beta \frac{\Delta  \varepsilon  }{2} \right)^2 \left[ 1 - \tanh^2 \left(  \beta \frac{ \Delta  \varepsilon  }{2} \right)\right].
\end{eqnarray}    
Independently of the state degeneracy $N$, this heat capacity has a maximum at $\beta \Delta  \varepsilon / 2 \simeq 1.2$, leading 
approximately to $T \simeq 0.417 \Delta \varepsilon$. This formula can give grossly indications on the possible appearance 
of the lowest peak in the low temperature region. For instance, from Fig. \ref{Fig:sps}, we can deduce that $\Delta \varepsilon_n \simeq
0.26$ MeV (Yb nuclei with RMF-PK1) leading to an expected bump around $0.1$ MeV, that is consistent with 
the bump observed in upper panel of Fig. \ref{fig:Ybcv} which appears at  0.07-0.08 MeV.

\section{Comparison with experiment}

\begin{figure}[htbp] 
\includegraphics[width=9cm]{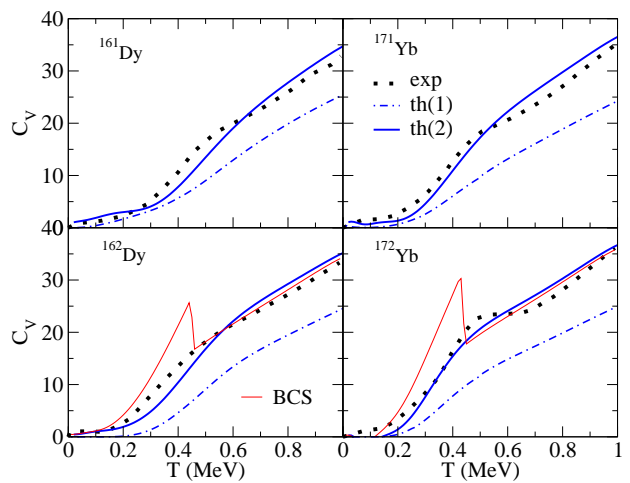} 
\caption{ (color online) Comparison between the calculated specific heats
and the specific heats extracted from experimental data \cite{Gut00,Sch01}.
By "th(1)" are indicated the results obtained with the level density generated
by the single-particle spectrum of RMF. The results "th(2)" are obtained 
employing in Eq. (\ref{extended-partition}) the partition function 
$\ln Z'_{nint}$ generated by the back shifted Fermi gas model. For comparison, 
we show also the results obtained within the FT-BCS approximation .}
\label{fig:expth} 
\end{figure}

The specific heats for Dy and Yb isotopes have been evaluated from the experimental
level density in Refs \cite{Gut00,Sch01}. How these results compare with the present calculations,
in the case of s.p. energies generated by RMF can be seen in Fig. \ref{fig:expth}.  This figure shows that
the calculated heat capacities underestimate significantly the experimental results.
 At first glance,   one might
conclude that the theory cannot quantitatively describe the experiments.
However, it should be kept in mind that in Refs. \cite{Gut00,Sch01} the heat 
capacities are obtained from the level density using specific assumptions. 
More precisely, in Refs.  \cite{Gut00,Sch01} it is used the experimental level density
for excitations below 8 MeV/nucleon while for higher excitation energies it
is employed the level density provided by the  Back Shifted Fermi Gas Model (BSFGM)  
(Eq. (5) of \cite{Sch01}). In the present calculations, we have used for all excitations
energies the level density generated by a discrete set of single-particle states.
This assumption is expected to work reasonable well for low energy excitations
but not for high energy excitations for which the contribution of the continuum
become important. Thus the underestimation of the $C_V$ at high temperatures by the
theoretical calculations appears to be related to the underestimation of the 
level density compared to the BSFGM.

\begin{figure}[htbp] 
% \vspace{0.8cm}
\includegraphics[width=9cm]{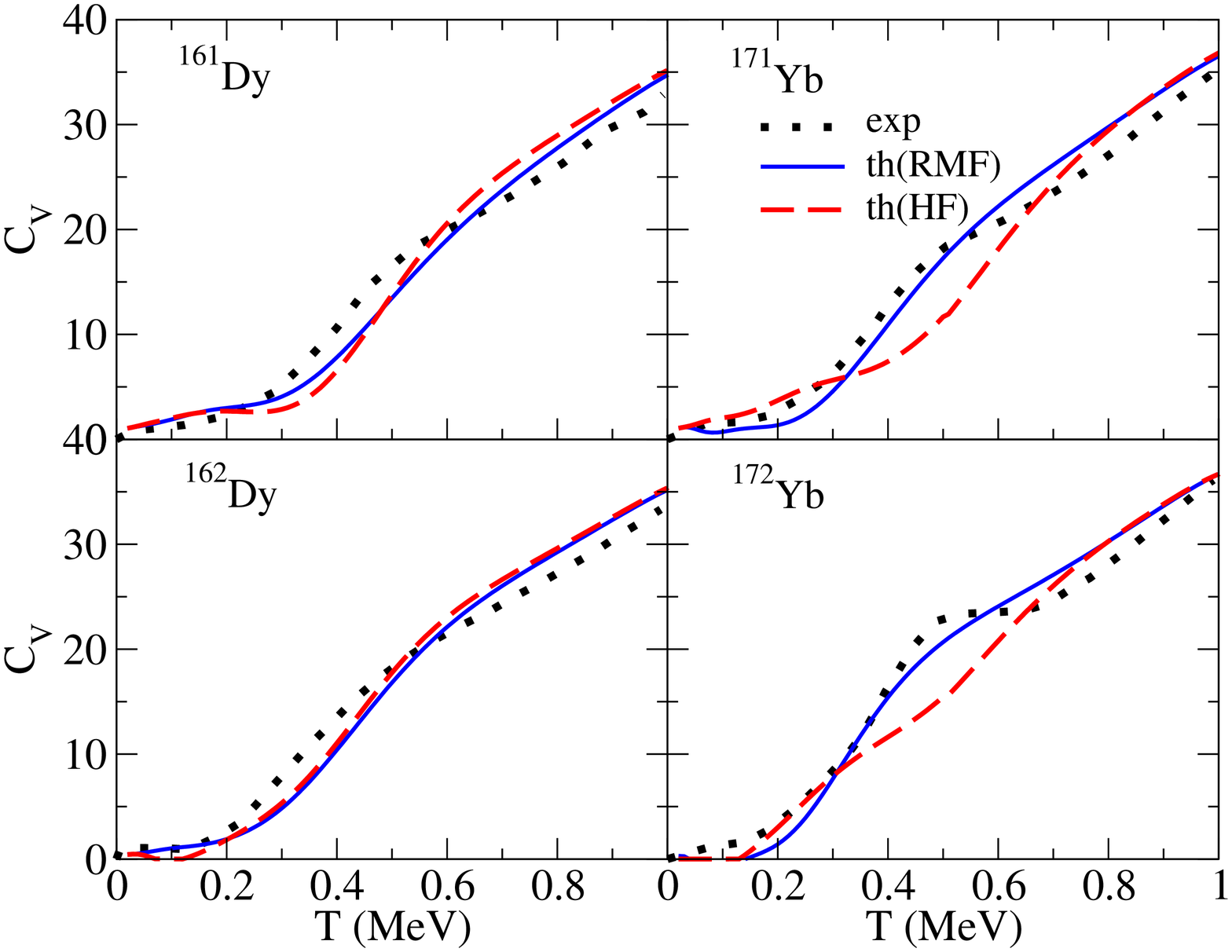} 
\caption{ (color online) Comparison between the specific heats
obtained  with the HF and RMF mean fields and with the extended
partition function (\ref{extended-partition}) calculated with
$\ln Z'_{nint}$ generated by the level density of  back shifted 
Fermi gas model.}
\label{fig:expth2} 
\end{figure}

To check that the difference between theory and experiment originates mainly from
the different treatment of level densities at  high excitation energies, where 
the pairing interaction is expected to not contribute, we have calculated the
specific heat with the partition function Eq. (\ref{extended-partition})
obtained using  $Z_{nint}$ evaluated with  BSFGM. 
The results are shown in Fig. \ref{fig:expth}. For comparison, for even
isotopes we show  also the specific heats obtained in the FT-BCS approach,
calculated  with $Z_{nint}$ generated by BSFGM, which present the unphysical
sharp transition between the superfluid and the normal phase. It can be observed
that the agreement of the FT-VAP results with the experiments is greatly
improved. In particular, the calculated specific heat is now joining the
experimental results at high temperatures. Globally, the S-shape behaviour of
the specific heat seems to be more pronounced than in the calculations. However,
we should keep in mind that, as we have discussed above, the shape of the
calculated specific heat is rather sensitive to the assumption made for the mean
field and to the strength of the interaction. This can be further seen in Fig. \ref{fig:expth2}, 
where the calculated specific heats obtained by using the RMF  and HF single-particle
levels  are compared to the experimental data. It can be noticed that the larger
differences are found for the Yb isotopes. These differences in the calculated specific 
heat are generated mainly by the different proton level density around the Fermi level
predicted by the  HF and RMF calculations (see Fig. \ref{Fig:sps}).  

It should be mentioned that a 
conclusion on the validity of RMF compared to HF would be erroneous. Indeed, changing the Skyrme functional in HF 
and/or the effective interaction in RMF
would completely change the level scheme and the comparison with experiments.  
 
\section{Conclusion}
 
In the present paper we have studied the effect of pairing correlations on specific 
heat of hot nuclei in the framework of a variation after particle-number projection
BCS formalism extended to finite temperature (FT-VAP). The calculation are done
in canonical ensemble and with  a Hamiltonian composed of a single-particle term, 
generated by Skyrme-HF or RMF calculations, and a pairing interaction of seniority type. 
The  pairing interaction is considered active in a limited window around the Fermi level. 
The contribution of the states from the pairing window to the canonical partition 
function is calculated in the FT-VAP approach and by direct diagonalization, the latter
being used to test the accuracy of FT-VAP results. The contribution of the states outside the 
pairing window is taken into account through the grand canonical partition function of 
a non-interacting system, which is reduced to the canonical representation
by the saddle point approximation. With this calculation scheme we have
evaluated the specific heat in the isotopes $^{161,162}$Dy and $^{171,172}$Yb.
It is thus shown that the pairing correlations have a significant influence on
the specific heat, especially for temperatures below $T=0.5$ MeV. The comparison
with the non-interacting systems shows clearly that the pairing correlations
contributes to the S-shape behaviour of the specific heat, as noticed in the
data extracted from experiment.

Compared to the specific heat extracted from experimental data, 
the calculations predicts much smaller values at larger temperatures.
It is shown that the FT-VAP approach is in fact able to predict 
results close to the experiment  if the contribution of the high 
energy excitations  is calculated not with the single-particle 
states provided by the mean field models but with the Back Shifted
Fermi Gas model, as actually done when the experimental specific 
heat is evaluated. The necessity to replace the non interacting partition
function at large excitation energy by the BSFGM one, clearly points out the 
important role of continuum part of the spectrum, neglected in these calculations.
The calculated specific heat  can be also influenced by the low-lying collective
states which are not accounted for by the Hamiltonian (1). These two issues
will be analysed in a future study.

\vskip 0.2cm
\noindent
{\bf Acknowledgments}
\vskip 0.1cm
This work was supported by the French-Romanian IN2P3-IFIN agreement and  by 
the Romanian Ministry of Education and Research through the grant Idei nr 57.
\appendix 

\section{Canonical partition function for  non-interacting particles}
\label{app:gc}   
We consider a system formed by a finite number of fermions distributed
in a set of single-particle states of energy $\epsilon_i$, generated,
for instance, by a self-consistent HF calculation. If there are no
residual interaction between the particles, the partition function
in the gran canonical ensemble can be written as
 \begin{eqnarray}
 \label{ZZ}
 \ln Z^{GC}_{nint} (\beta,\mu) = \sum_{i}  \ln \left[1 +
 e^{-\beta(\epsilon_{i}-\mu)} \right]  
 \;,
 \end{eqnarray}
where $\mu$ is the chemical potential. 

The  grand-canonical partition function can be used to obtain
a simple approximation for the partition function in the
canonical ensemble, expressed in terms of single-particle quantities. 
This can be done by  applying the saddle point approximation. 
One thus gets the following approximation for the canonical 
partition function \cite{Alh03}
\begin{eqnarray}
\label{partition-N}
\ln Z' \approx \ln Z^{\rm GC} +\beta E_0 -\beta\mu N -{1\over 2}\ln \left( 2 
\pi   \langle(\Delta N)^2\rangle \right)
\;.
\end{eqnarray}
The chemical potential $\mu$, the particle number $N$ and its fluctuation $\Delta N$
are given by the  equations
\begin{eqnarray}\label{particle-number}
N = \sum_i f_{i}, ~{\rm and}  ~\langle(\Delta N)^2\rangle = \sum_{i} f_{i}(1-f_{i}) 
\end{eqnarray}
where $f_i=[1+e^{\beta(\epsilon_i-\mu)}]$ is the Fermi-Dirac distribution.
From the canonical partition function one can calculate the specific heat by
doing numerical derivatives or using the formulas (B4, C1-C3) of Ref \cite{Alh03}.

\end{document}